\documentclass[letterpaper]{article}
\usepackage{iccc}
\usepackage{graphicx}
\usepackage{graphics}
\usepackage{lipsum}

\usepackage{times}
\usepackage{helvet}
\usepackage{courier}
\title{Interpolating GANs to Scaffold Autotelic Creativity}
\author{Ziv Epstein, Oc\'{e}ane Boulais, Skylar Gordon, Matt Groh \\ \newline \texttt{\{zive, oceane, sfgordon, groh\}@mit.edu} \\MIT Media Lab}

\begin{document} 
\maketitle
\begin{abstract}
\begin{quote}




The latent space modeled by generative adversarial networks (GANs) represents a large possibility space. By interpolating categories generated by GANs, it is possible to create novel hybrid images.
We present ``Meet the Ganimals," a casual creator built on interpolations of BigGAN that can generate novel, hybrid animals called ganimals by efficiently searching this possibility space. Like traditional casual creators, the system supports a simple creative flow that encourages rapid exploration of the possibility space. Users can discover new ganimals, create their own, and share their reactions to aesthetic, emotional, and morphological characteristics of the ganimals. As users provide input to the system, the system adapts and changes the distribution of categories upon which ganimals are generated. As one of the first GAN-based casual creators, Meet the Ganimals is an example how casual creators can leverage human curation and citizen science to discover novel artifacts within a large  possibility space.

\end{quote}
\end{abstract}

\section{Introduction}
Generative adversarial networks (GANs) \cite{goodfellow2014generative} are a subclass of generative models that enable anyone to generate photo-realistic images with a single click of a button \cite{karras2019style}. Some have heralded this innovation as the end of design, whereby machine intelligence will replace human creation \cite{slate}. However, a larger contingent considers these new generative technologies as yet another tool in an artist's toolkit, which offers new expression with its novel affordances \cite{hertzmann2018can}. Since the application of generative models does not require formal artistic training or technical expertise, these models can serve as scaffolding to generate large possibility spaces that can be embedded in casual creator systems \cite{compton2015casual}. Recent platforms such as RunwayML, GANBreeder, GANPaint, and DeepAngel have already started to use the new medium of GANs for casual creation.  

The key challenge for GAN-based casual creators is designing systems that ``supports a state of creative flow'' - whereby users and the generative models can co-create new artifacts in a collaborative, coordinated and organic dialogue, towards the idea of \textit{mixed-initiative co-creativity}.~\cite{yannakakis2014mixed,acharya2019building}. 
\begin{figure}[h]
    \centering
    \includegraphics[width=0.5\textwidth]{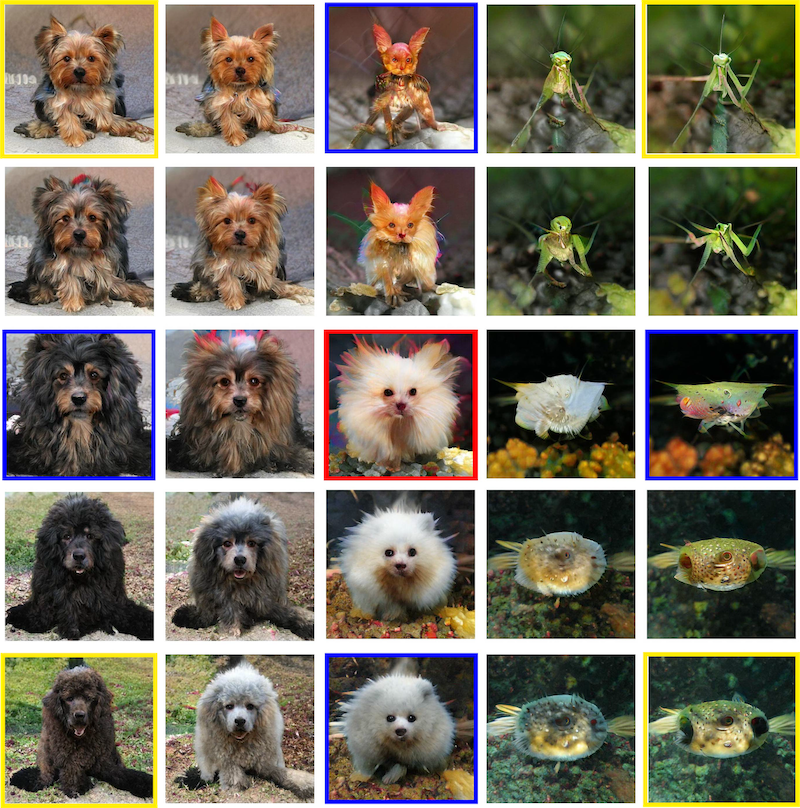}
    \caption{Schematic for interpolation. The yellow images in the four corners represent the praying mantis, a boston terrier, a pufferfish and poodle categories respectively, which we call generation zero ($G_0$) ganimals, the images in the four outer mid-points in blue are hybrids of two $G_0$ ganimals which we call $G_1$ ganimals,  and the center image is a $G_2$ ganimal, which is a combination of all four $G_0$ ganimals.}
    \label{fig:my_label}
\end{figure}

We introduce one such casual creator system, Meet the Ganimals, that allows users to selectively create new artificial hybrid species by interpolating between categories modeled by BigGAN \cite{brock2018large}. Trained on images with 1000 categorical labels, BigGAN embeds each category in a high-dimensional latent space. This space can be smoothly traversed such that images of mixed categories can be synthesized via interpolating the categories. Figure~\ref{fig:my_label} presents examples of images generated from single and mixed categories. The original BigGAN model was trained on 1000 categories, and we restrict BigGAN to the 396 animal categories. The goal of constraining the model to animal categories is to focus the experience on discovering and breeding hybrid animals –- what we call ganimals. 

\begin{figure*}[h]
    \centering
    \includegraphics[width=\textwidth]{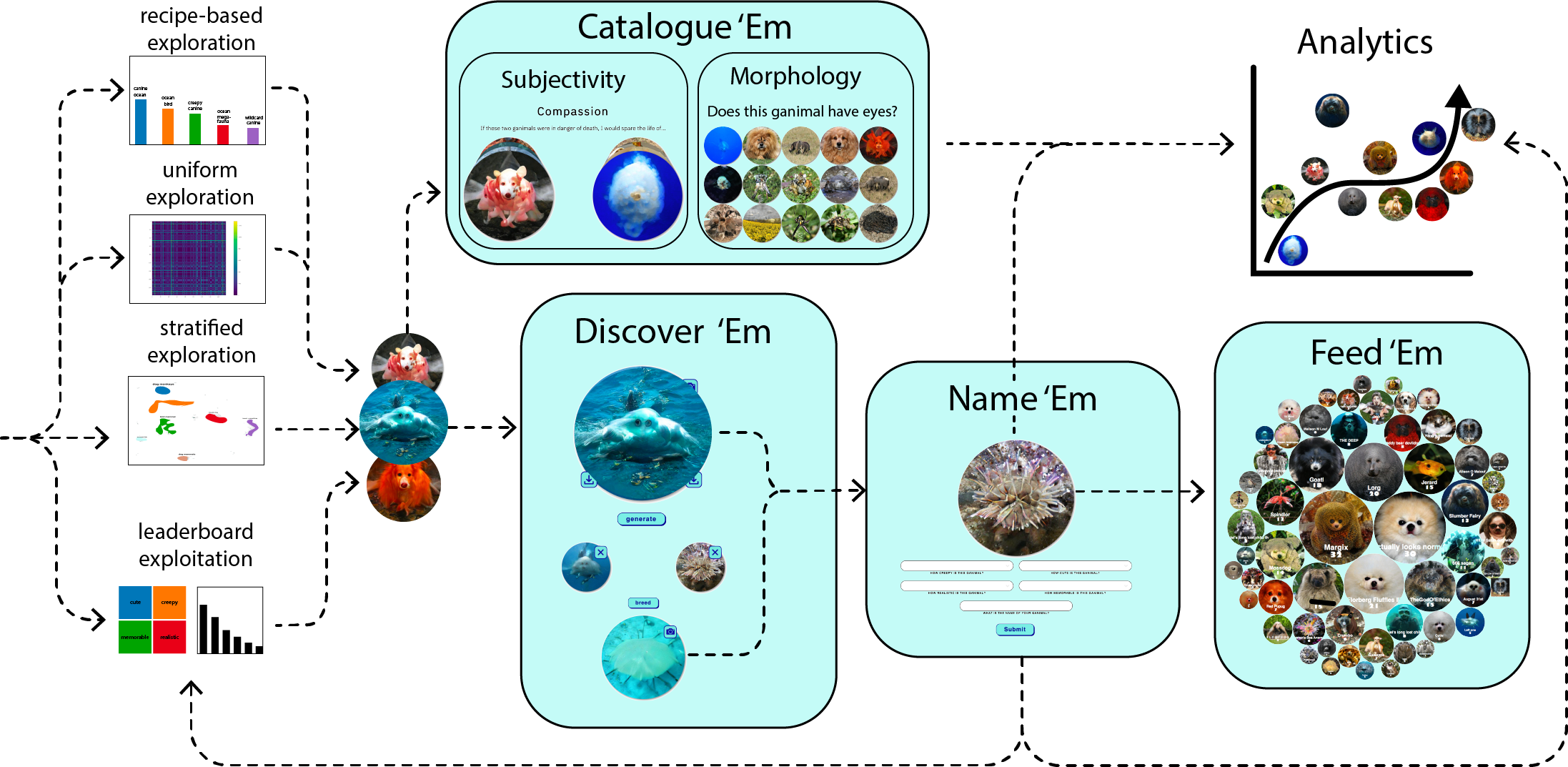}
    \caption{System map of the Meet the Ganimals Platform }
    \label{fig:map}
\end{figure*}

Unlike professional creative systems which provide users with many precise tools to craft artifacts directly, Meet the Ganimals is a simple interface designed to promote the exploration of a vast possibility space. The possibility space includes three generations of ganimals; 396 $G_0$ ganimals that correspond directly to the ImageNet animal categories, 78,210 $G_1$ ganimals from hybrid pairs of $G_0$ ganimals, and 3,058,362,945 $G_2$ ganimals come from hybrid quadruples of $G_0$ ganimals. The possibility space is even larger when accounting for variations that come from truncation inputs and random seeds. 

With such a large possibility space, it is difficult to find the near-superlative ganimals e.g. the cutest, creepiest, most memorable ganimals. Meet the Ganimals confronts this seemingly intractable search problem with two innovations. First, Meet the Ganimals simplifies exploration into a two-part creation interface: Users generate a large number of $G_1$ ganimals to find the ones they like best, and then they breed the chosen $G_1$ ganimals into $G_2$ ganimals. Second, instead of randomly combining categories to generate $G_1$ ganimals in this first stage, the system instead balances exploring new permutations, and exploiting previously popular permutations, as indicated from crowd signals from other users. These innovations build on earlier interfaces where user-generated landmarks serve as navigation elements in a large parametrically-defined possibility space \cite{talton2008collaborative,harding2018biomorpher}.  


\section{Related Work}
Several recently developed platforms explore how GANs serve as scaffolding for autotelic creativity. RunwayML provides a simple interface such that non-technical artists can use state-of-the-art neural network models (e.g. style transfer and super resolution) in their work. GANPaint is a scene drawing tool that allows users to add or remove trees, grass, and other natural features with a simple click \cite{bau2018gan}. DeepAngel is an online tool that removes objects from images users upload via masking and generative inpainting \cite{groh2019human}. While such functionality exists in Photoshop (e.g. content-aware fill), DeepAngel provides a one-click interface, sidestepping the technical skills required for photo-editing.

Other related platforms have explored creating collaborative media via crowd signals and genetic algorithms. The R/Place experiment on Reddit allowed users to collectively recolor pixels of a dynamically changing image \cite{rappaz2018latent}. Electric Sheep generates procedural animations using crowd signals in an evolutionary algorithm \cite{draves2005electric}. PicBreeder demonstrated how pictures could be evolved collaboratively to rapidly explore a possibility space and proliferate fascinating discoveries while promoting individual exploration \cite{secretan2008picbreeder,secretan2011picbreeder}.

ArtBreeder is an example of a casual creator built on interpolating GANs \cite{simon2019ganbreeder}. From the perspective of a user, ArtBreeder is a platform for creating new artworks by blending existing images.

Meet the Ganimals builds on these projects to combine GAN scaffolding with collective feedback while focusing on the domain of hybrid animals.

\section{System Overview}
Meet the Ganimals is designed with modern UI/UX paradigms built on BigGAN to serve as a mixed-initiative co-creation tool whereby users are both creators and consumers of the possibility space. From April 20th to May 20th, 51,110 ganimals were generated, and 10,587 ganimals were bred by 4,392 users. The system map for the platform is shown in Figure~\ref{fig:map}. 

\subsection{Random Stimulus for Exploring Possibility Space} 
In \textit{The Book of Imaginary Beings}, Jorge Luis Borges
wrote about his compilation of created creatures, saying that ``the book $\cdots$ is not meant to be read straight through; rather, we should like the reader to dip into these
pages at random, just as one plays with the shifting patterns
of a kaleidoscope'' \cite{borges2002book}. Echoing Borges, this system leverages the idea of the random stimulus principle of lateral thinking to offer a stochastic exploration of the possibility space \cite{beaney2005imagination}. In the ``Discover 'Em'' page, users are shown ganimals randomly generated using a bandit algorithm that balances exploration of an unseen possibility space with the popularity of the discovered space. In particular, $G_1$ ganimals are generated and presented to users according one of four selection procedures: (1) 30\% of the time that ganimal is generated using a carefully feature engineered stochastic process that samples pairs that we as designers found to be compelling (``recipe-based exploration''), (2) 30\% of the time by random uniformly sampling two animal categories to breed (``uniform exploration''), (3) 30\% of the time by randomly sampling two animal categories to breed stratified by species (``stratified exploration''), and (4) 10\% of the time by sampling a ganimal from the top rated ganimals, proportional to its order in the leaderboard for a random one of the following characteristics: cute, creepy, realistic, or memorable (``leaderboard exploitation''). 

For the recipe-based exploration, we carefully curated an ad-hoc generative process that we found created high-quality ganimals. In particular, we defined five cores - sets of conceptually similar ImageNet categories that are well-suited for blending for aquatic, canine, bird, megafauna, and wildcard categories. We then randomly blend these cores in order to create diversity in the resulting ganimals. 

For stratified exploration, we uniformly sample a pair of animal species and sample an ImageNet category that corresponds to that species. For the majority of categories, there is a one-to-one correspondence between ImageNet categories and species. However, there are 118 categories of dogs (Canis lupus familiaris). The stratified exploration downsamples the frequency of dogs relative to the frequency in which dogs appear in ImageNet categories to promote diversity in the kinds of ganimals created. Users can curate $G_1$ ganimals and blend $G_1$ ganimals to create their own $G_2$ ganimal, which can be named and given its own unique hyperlink. This process supports the creative flow that allows users to efficiently explore the system's possibility space, view a diverse array of combinations, and add their own creativity to the that of the system to create their own artifacts.  

\begin{figure}
    \centering
    \includegraphics[width=0.5\textwidth]{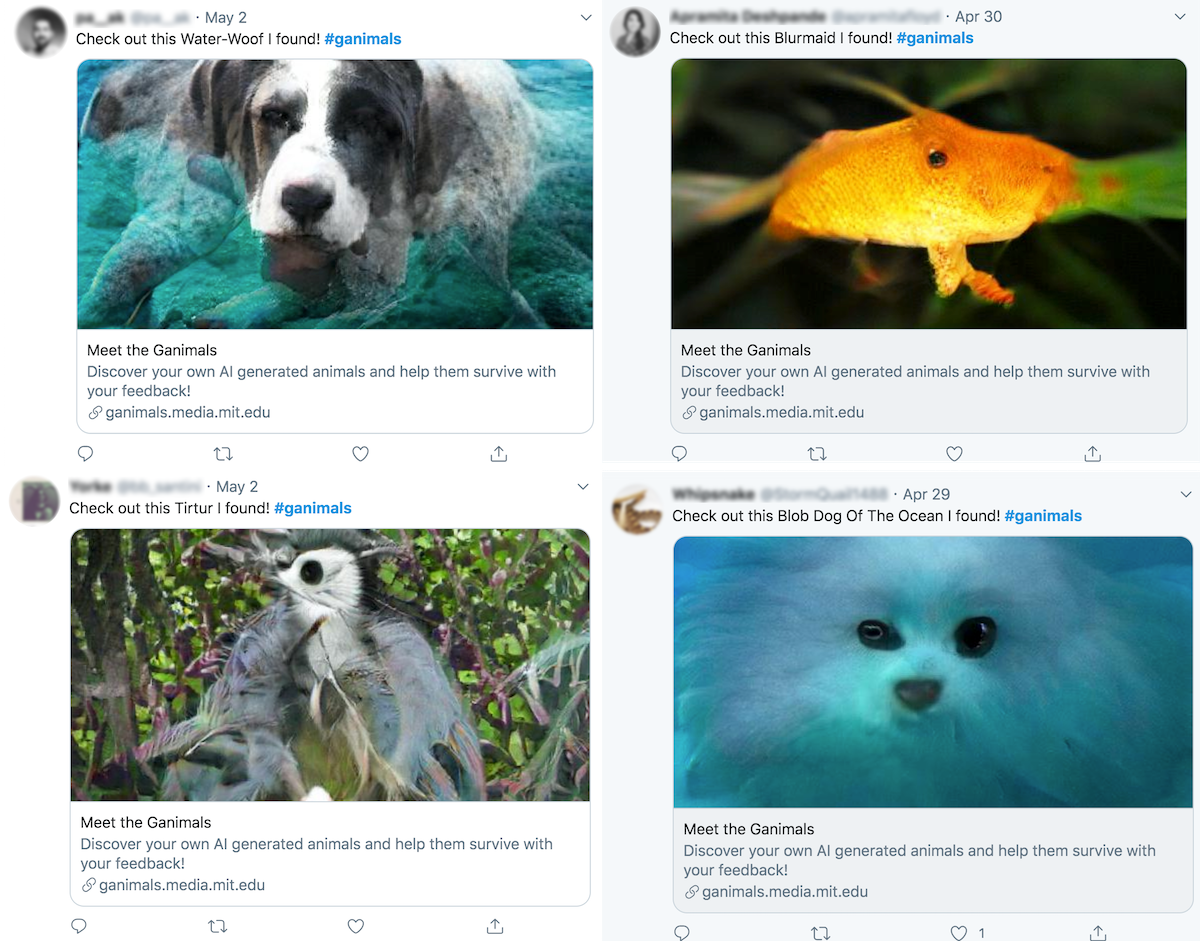}
    \caption{Selected screenshots of tweets from (anonymized) users sharing the ganimals they have discovered and named.}
    \label{fig:tweets}
\end{figure}

Users feel a sense of pride and ownership over the ganimals they create, which they have shown by sharing discoveries on social media  (see Figure~\ref{fig:tweets}).

\subsection{Towards a Citizen Science of GAN Subjectivity}
From identifying exoplanets \cite{zink2019catalog} and Christmas birds \cite{birds} to detecting changes in climate \cite{scistarter} and coral reef coverage \cite{raoult2016gopros}, citizen science projects have been core part of engaging the public in global scientific operations.  Ultimately, participants engage in these projects because it is of individual interest to participate in a public (and therefore social) network \cite{lukyanenko2019citizen}, and because they are aesthetically pleasing and easy-to-use \cite{bonney2009citizen}.  

Meet the Ganimals has no metric for success, efficiency, or productivity, and there is no way for a user to demonstrate technical artistic or design skills. Instead, users are motivated by naming privileges for unseen ganimals, and general curiosity. As such, Meet the Ganimals is well suited for casual citizen science.

In  the ``Catalogue 'Em'' page, users have the opportunity to take on the role of citizen scientists (i.e. ``casual'' scientists) to help answer scientific questions about the ganimals: Do ganimals with canine morphological features look cuter than the rest \cite{kaminski2019evolution}? Does curation decrease as the underlying animal categories diverge in evolutionary time \cite{miralles2019empathy}? Do descendents of charismatic megafauna emerge as the most popular \cite{bennett2017conservation}? 

To explore such questions, users can recount their subjective and emotional perspectives of the ganimals, as well as annotate their morphological features. For morphology features, users can annotate whether or not the ganimals have a head, eyes, a mouth, a nose, legs, hair, scales, feathers, live underwater, or are bigger than a house cat. For subjective perspectives, users can annotate how much compassion and empathy they feel towards the ganimals, as well as how cute, memorable, realistic and creepy they are.  

\begin{figure}
    \centering
    \includegraphics[width=0.5\textwidth]{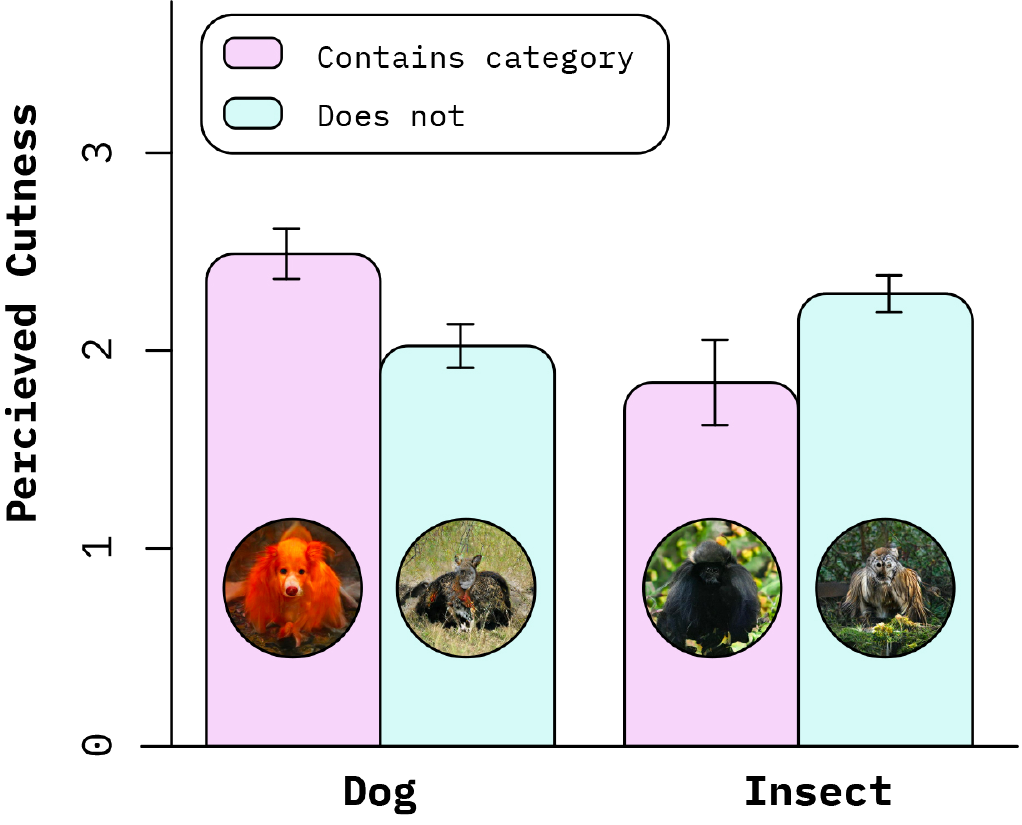}
    \caption{Perceived cuteness rating of ganimals with or without specific categories. Images are exemplary ganimals from that set.}
    \label{fig:ploz}
\end{figure}

This process allows for a deeper understanding of how animal morphologies relate to subjective perception. In particular, we can correlate subjective evaluations of ganimals with their other characteristics - such as crowd annotated morphology features, or the number of ``dog'' categories present within that ganimal. As a preliminary analysis, we find that ganimals contain at least one dog are statistically significantly cuter than those that do not, and that ganimals that contain at least one insect are statistically significantly less cute than those that do not (see Figure~\ref{fig:ploz}). Such knowledge can inform the design of future generative algorithms that use crowdsourced labels to surface maximally compelling artifacts. 

When users take on the role of an explorer in the realm of ganimals, they not only become a creator, but also take on a role as a participant in the scheme of a broader investigation.  By departing from the traditional role of a data-driven "scientist" who seeks a conclusion to a hypothesis, Meet the Ganimals empowers the exploration-driven creator: casual citizen scientists do not participant in the context of accomplishing a specific task or aim, and largely are not driven by a particular hypothesis.  This unique dynamic thereby can leverage discovery to collect in-the-field results in a host of different ecosystems.

\section{Random World Assignment to Explore Local Ecologies}
\par In line with the Music Lab experiment \cite{salganik2006experimental}, a final ingredient for the Meet the Ganimals system is the random assignment of each participant to a ``world,'' with its own local ecology that involves independently from those of other worlds. Each world is initialized with a fixed ``seed set'' of 100 ganimals, randomly selected using the random stimulus approach discussed above.  Then users assigned to a particular world interact with this seed set plus the ganimals discovered and bred by users assigned to that world.  In the ``Feed Em'' page, users can feed the ganimals they like the best. Well-fed ganimals are promoted and remain in this view, while unfed ganimals disappear. In addition, the fourth selection procedure from the bandit algorithm (``leaderboard exploitation'') only pulls ganimals from the world corresponding to that user.

The design of the ``Feed Em'' page can differ across worlds, allowing cross-world comparisons to serve as an A/B test for how different UX/UI patterns affect emergent ecologies. For example, one might ask how the layout of the ``Feed Em'' page (e.g. a linear feed-like view versus a more spatial ecological view) changes the resulting diversity of the ganimals in that world.\footnote{A deep dive into the experimental design and measurement approach of such a research question is beyond the scope of this paper, which focuses on the overview of the casual creator itself. However interested readers can learn more by reading the pre-analysis plan here:  https://aspredicted.org/65nv7.pdf}. The random assignment of users to different worlds that evolve independently provides a virtual laboratory to compare  behavior and curation across worlds, and causally assess the impact of design interventions.

\section{Discussion}

GAN architectures force two computational agents, the generator and discriminator, to compete against each other with the goal of creating a statistical model resembling the training data. Casual creators built on GAN architectures introduce a third agent, a casual human collaborator, into the loop to explore the most intriguing parts of latent space.

With a simple interface for creating and curating images of hybrid AI generated animals, users are motivated to engage in computational creativity for no other reasons than their own curiosity and the chance to name their creations and discoveries. The autotelic motivation drives the interactions within the casual creator and as a result, the system provides insights into what intrigues people \cite{yannakakis2014mixed}. 


 In many creative endeavors, the production and consumption of artifacts are separated, which can lead to undifferentiated production and passive consumption. Human-in-the-loop casual creators built upon GANs are a new medium that blends production and consumption of media into a singular creative process. While Meet the Ganimals focuses on generating images of hybrid animals in particular, it is but one of a growing number of casual creators built on interpolating the GAN latent space of other cultural artifacts. Beyond animals, artifacts as varied as facial expressions, architectural landmarks and fashion are emerging domains where GANs could serve as scaffolding for casual creators \cite{zhu2020domain}, paving the way for new forms of human-AI collaboration. With a well-constrained casual creator, the frontiers of 
 the GAN latent space are within reach. 
 

\bibliographystyle{iccc}
\bibliography{iccc}


\end{document}